\documentclass[twocolumn,aps,superscriptaddress,showpacs]{revtex4}
\usepackage[dvips]{graphicx}
\usepackage[dvips]{rotating}
\usepackage{dcolumn}\usepackage{lscape}
\usepackage{float}\usepackage{subfigure}
\usepackage{fancybox}

\begin{document}

\title{Power-law distribution of individual Hirsch indices, 
the comparison of merits in different fields, and the relation
to a Pareto distribution}
\author{Pekka Pyykk\"o}
\affiliation{Department of Chemistry, University of Helsinki, P.O. Box 
55 (A.I. Virtasen aukio 1), FI-00014 Helsinki, Finland\\
E-mail: Pekka.Pyykko@helsinki.fi}

\begin{abstract}
A data set of Hirsch indices, $h$, for Finnish scientists in certain fields
is statistically analyzed and fitted to $h(n) =Pn^p$
for the $n$-th most-quoted scientist.
The precoefficient $P$ is characteristic for the field
and the exponent $p$ is about -0.2 for all data sets considered. 
For Physics, Chemistry and Chemical Engineering, the $P$ are
49.7(8), 41.3(6), and 21.4(6), respectively. These $p$ values
correspond to  Pareto exponents of about -7 for the
distribution of Hirsch indices $h$. 
\end{abstract}

\maketitle

\section{Introduction}
The Hirsch index $h$ \cite{Hirsch:05,Hirsch:05b} provides a rough but robust
measure on the total citation impact of an individual, 
until the time of observation.
More exactly it means having $h$ papers, each cited at least $h$ times.
In addition to persons, it also can be defined for universities, journals etc.
The values are very different for different fields and the question
is, how to compare the values between fields?

We had available a small data set of the $h$ values in Chemistry, Physics,
and Chemical Engineering for Finnish scientists. A statistical study
reveals an interesting power-law distribution and gives a hint on
the relative weighting factors that may apply between different fields.

\section{Method and results}

\begin{figure}
\includegraphics[width=6cm,angle=-90,clip=]{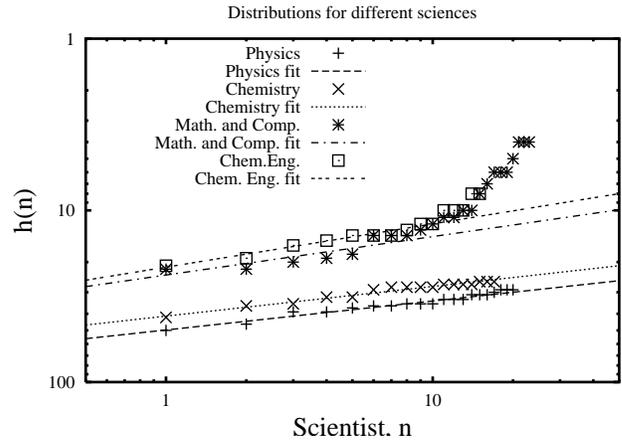}
\caption{The fits (1) for Physics, Chemistry, Mathematics plus Computer Science,
 and Chemical Engineering. For the two latter fields the entire data sets are
 shown as points, although the fits only include the $k$ highest points in
Table \ref{tab:res}.}
\label{fig:1}
\end{figure}

The data were determined from the ISI Web of Knowledge using the data set
in General Search from 1945 onwards. 
This database only contains references in journals to papers in
journals. 
Most data points were obtained in November 2005. 
The most-quoted one-third of the points inside each area, $k$, was fitted
using Gnuplot to 
\begin{equation}
h(n) = P n^p
\label{eq:fit}
\end{equation}
where $h(n)$ is the $h$ of the $n$:th-most quoted scientist, 
$P$ is a precoefficient and $p$ is an exponent, found to be surprisingly 
constant between different fields. The obtained values are shown in Table
\ref{tab:res} and the quality of the fits is demonstrated in Fig.
\ref{fig:1}. The figures in parentheses give the asymptotic standard error.
In this data set, for the given country at the given time,
the workers in  different areas mostly share the same background
and general working conditions, like the typical research-group size
and budget. Assuming that they also are 
equally gifted and hard-working,
we then suggest that the ratios of $P$ between different fields
would form a possible basis for comparing scientific merit between fields.

Podlubny\cite{Podlubny:05} recently compared the total numbers of
citations in various fields in United States. 
He found them to be fairly constant from
1992 to 2001 and suggested that they would form a useful normalization
factor for comparing individual scientific performance between fields.

\begin{table}
\caption{The fits for certain areas. $k$ is the number of points included
in the fit. All data refer to Finland.}
\begin{ruledtabular}
\begin{tabular}{lrlll}
Area & $k$ & $P$ & $p$ &  \\
\hline
Medicine & 4 & 90(3) & -0.22(3) & \\
Bio/eco  & 5 & 59(4) & -0.23(7)& \\
Physics  & 14& 49.7(8) & -0.169(9) & \\
Chemistry& 17& 41.3(6) & -0.173(7) & \\
Math and Comp & 8 & 23.8(1.5) & -0.22(5) & \\
Chem. Eng. & 5 & 21.4(6) & -0.25(3) & \\
\end{tabular}
\end{ruledtabular}
\label{tab:res}
\end{table}

In Table \ref{tab:Pod} we compare the present relative $P_{\rm rel}$
factors (with Physics normalized to 1) to the square roots of Podlubny's
relative citation numbers. An average of his 1992-2001 data is used.

Recall here that the lower limit for the total number of citations, 
$N_{c,tot} = h^2$ and a typical number is\cite{Hirsch:05b}
\begin{equation}
N_{c,tot} = ah^2,
\label{eq:Ntot}
\end{equation}
with $a$ about 3-5 \cite{Hirsch:05b}.
\begin{table}
\caption{The relative prefactors, $P_{\rm rel}$, with Physics normalized 
to one and the square roots of the number of total citations, 
$(C_{\rm rel})^{1/2}$, with Physics normalized to one.}
\begin{ruledtabular}
\begin{tabular}{lrr}
Area &  $P_{\rm rel}$ &$(C_{\rm rel})^{1/2}$  \\
\hline
Medicine & 1.8 & 2.0  \\
Bio/Eco  & 1.2 & \\
Physics  & 1 & 1  \\
Chemistry& 0.83 & 0.88 \\
Math and Comp & 0.48 & 0.23  \\
Chem. Eng. & 0.43 &  \\
\end{tabular}
\end{ruledtabular}
\label{tab:Pod}
\end{table}
\section{Further data sets}
\begin{figure}[H]
\includegraphics[width=6cm,angle=-90,clip=]{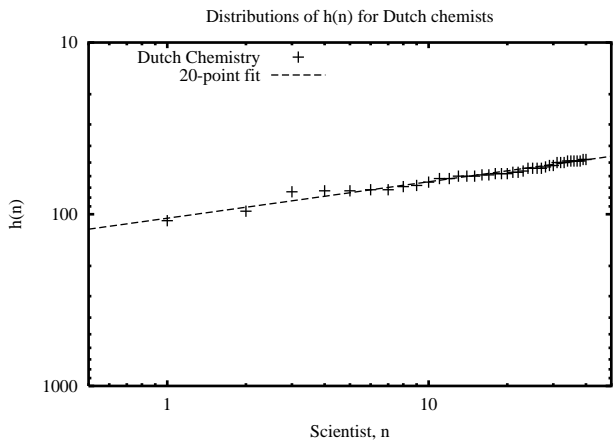}
\caption{The $h$ values of forty 'Dutch' chemists from Faas\cite{Faas:06}.
The line is fitted to the points 1-20 and has $p$ = -0.212(11).}
\label{FaasFig}
\end{figure}

A list of the $h$ values for 40 'Dutch' chemists was published by
Faas\cite{Faas:06}. Both people of Dutch origin anywhere in the World,
and people from anywhere, working in The Netherlands were included.
As seen from Fig.\ref{FaasFig}, all points
fit well the values $P$ = 105.5(2.4), $p$ = -0.212(11).
\section{Relation to the Pareto distribution}
In economic theory, V. Pareto 
found in 1896 \cite{Pareto:96} 
the number of holders of  income $I$ in a country to scale
for high incomes as $I^x$, with $x$ about -2
(\cite{Pareto:96},  see ref. \cite{Farmer:05,Newman:05}).
The same law was
found by Zipf to hold for word frequencies in linguistics and by Lotka
for numbers of papers among authors\cite{Lotka:26}. It is
known in many other fields, like size distributions of cities in a
country, earthquakes, wars etc. \cite{Newman:05}.

From eq. (\ref{eq:fit}), 
\begin{equation}
n(h) = (h/P)^{1/p}.
\label{eq:inv}
\end{equation}
Introducing the density of individuals per unit of $h$, $N(h)$,
\begin{equation}
n = \int_{h_n}^\infty N(h') dh',
\label{eq:cum}
\end{equation}
we can interprete $N(h)$ as the derivative
\begin{equation}
N = -dn/dh.
\label{eq:interp}
\end{equation}
Then, using eq.\ref{eq:inv},
\begin{equation}
N(h) =  P^{-1/p} h^{\frac{1}{p}-1}.   
\label{eq:Par}
\end{equation}
For the Finnish $p$ for Physics and Chemistry, the corresponding Pareto
exponent $x$  would become -6.9 and -6.8, respectively. 

The main conclusions are that the $P$ value for Chemical Engineering is
about half of that for Chemistry, and that the $p$ values for the data sets
considered are about -0.2.
\begin{acknowledgments}
Claus Montonen pointed out ref. \cite{Podlubny:05}.
\end{acknowledgments}

\newpage


\begin{thebibliography}{99}
\bibitem{Hirsch:05} J.~E. ~Hirsch, arXiv:physics/0508025 v5, 29 Sep 2005.
\bibitem{Hirsch:05b} J.~E. ~Hirsch, Proc. Natl. Acad. Sci. (USA) {\bf 102},
16569 (2005).

\bibitem{Podlubny:05}  I. ~Podlubny, J. Scientometrics {\bf 64}, 95 (2005).
arXiv math ST/0410574.

\bibitem{Faas:06} F. ~Faas, Chemisch 2 Weekblad (March 11, 2006), p. 18.

\bibitem{Pareto:96} V. ~Pareto, {\em Cours d'Economie Politique},
Droz, Gen{\`e}ve (1896). As quoted in ref. \cite{Newman:05}.


\bibitem{Farmer:05} J.~D. ~Farmer, M.~Shubik, and E.~Smith, 
Physics Today {\bf 58}, No. 9, 37 (2005).

\bibitem{Newman:05} M.~E. ~J. ~Newman, Contemporary Physics {\bf 46}, 323
(2005).

\bibitem{Lotka:26} A. ~J. ~Lotka, J. Washington Acad. Sci., 
{\bf 16}, 317 (1926).


\end{thebibliography}
\end{document}